\documentclass[%
 reprint,
 nobalance,
 amsmath,amssymb,
 aps,
 showpacs, showkeys,
]{revtex4-2}

\usepackage{graphicx}
\usepackage{dcolumn}
\usepackage{bm}
\begin{document}

\title{Exact Density Profiles of 1D Quantum Fluids in the Thomas-Fermi Limit: \\Geometric Hierarchy to the Tonks-Girardeau Gas}

\author{Hiroki Suyari}
 \email{suyari@faculty.chiba-u.jp}
 \email{suyarilab@gmail.com}
 \thanks{ORCID: 0000-0003-2624-0902}
\affiliation{%
 Graduate School of Informatics, Chiba University, 1-33, Yayoi-cho, Inage-ku,
Chiba 263-8522, Japan\\
}%

\date{March 2, 2026}

\begin{abstract}
We present a geometric framework for 1D quantum fluids across interaction regimes in the Thomas-Fermi limit.
Based on the Linearization Principle via the $q$-logarithm, macroscopic density profiles form a discrete hierarchy: the ideal Bose gas ($q=1$), the mean-field Gross-Pitaevskii condensate ($q=-1$), and the strongly correlated Tonks-Girardeau gas ($q=-3$).
We further derive a universal sound velocity scaling, $c \propto \rho^{(1-q)/4}$, valid in the interacting regimes ($q \le -1$).
This establishes a non-perturbative link between static geometry and dynamical excitations in many-body systems.
\end{abstract}

                              
\maketitle


\section{Introduction}
\label{sec:introduction}

The Gross-Pitaevskii equation (GPE) is the standard mean-field model for Bose-Einstein condensates (BECs)~\cite{Pitaevskii2003}.
As a nonlinear Schr\"odinger equation, it successfully captures the macroscopic behavior of quantum fluids within the mean-field approximation, where the inter-particle interaction is effectively described by a cubic nonlinearity.
However, the validity of the standard GPE is limited to the weak-coupling regime.
In one-dimensional (1D) systems, as the effective interaction strength increases, the gas enters a strongly correlated regime where the mean-field approximation breaks down.
The extreme limit of infinite repulsion is the \textit{Tonks-Girardeau (TG) gas}, where bosons mimic the exclusion statistics of fermions due to impenetrability~\cite{Girardeau1960, Kinoshita2004}.
The density profile of the TG gas in a harmonic trap follows the Wigner semicircle law, 
which is fundamentally different from the inverted parabola profile (Thomas-Fermi approximation) of the standard BEC.

The exact microscopic description of 1D interacting bosons with contact interactions is provided by the Lieb-Liniger model~\cite{LiebLiniger1963}.
The interaction regime is characterized by the dimensionless parameter $\gamma = m_0 g/\hbar^2 \rho$, where $m_0$ is the particle mass, $g$ is the interaction strength, and $\rho$ is the 1D density.
While the Lieb-Liniger model is exactly solvable via the Bethe ansatz for a homogeneous system, extracting the macroscopic density profile $n(x)$ in the presence of an external harmonic trap typically relies on the Local Density Approximation (LDA) combined with numerical solutions.
Conventionally, the weak and strong correlation limits---the mean-field BEC ($\gamma \ll 1$) and the TG gas ($\gamma \gg 1$)---are described by distinct mathematical frameworks: the nonlinear GPE for the former, and the Bose-Fermi mapping for the latter.
This raises the question of whether macroscopic density profiles across these disparate correlation regimes can be analytically captured within a single geometric framework, bypassing the need for numerical LDA.

In this Letter, we introduce a geometric framework, the \textit{Linearization Principle}, to describe these systems in a unified manner.
Within this framework, macroscopic density deformations arise from the geometric curvature of the state space.
By introducing the $q$-logarithm \cite{Tsallis1988, Tsallis2009} as a natural coordinate that linearizes the stationary state, we derive a family of analytical density profiles in the Thomas-Fermi limit.
This geometric framework reveals a discrete hierarchy governing quantum fluids.
We show that a single geometric index $q$ parameterizes the physical regime: $q=1$ yields the ideal Bose gas, $q=-1$ recovers the Thomas-Fermi profile of the GPE, and $q=-3$ yields the fermionized Wigner semicircle law of the TG gas.
Furthermore, we demonstrate that this geometric structure extends to dynamical excitations, yielding the sound velocity scaling $c \propto \rho^{(1-q)/4}$ across the interacting regimes ($q \le -1$).


\section{Theory}
\label{sec:theory}

\subsection{The Nonlinear Schr\"odinger Equation and the Thomas-Fermi Limit}
The starting point of our discussion is the standard Gross-Pitaevskii equation (GPE), which describes the ground state of a Bose-Einstein condensate in the mean-field approximation:
\begin{equation}
    i\hbar \frac{\partial \psi}{\partial t} = -\frac{\hbar^2}{2m_0} \nabla^2 \psi + V_{\text{ext}}(x) \psi + g |\psi|^2 \psi,
    \label{eq:GPE}
\end{equation}
where $g$ represents the interaction strength.
For a large number of particles with repulsive interactions, the interaction energy dominates over the kinetic energy (quantum pressure) associated with the spatial variations of the density.
Neglecting the kinetic energy term $-\frac{\hbar^2}{2m_0} \nabla^2 \psi$ leads to the Thomas-Fermi (TF) approximation.
Our goal is to derive the macroscopic density profile of this limit---and its generalizations for strong coupling regimes---from a geometric principle.

\subsection{The Linearization Principle}
We introduce the \textit{Linearization Principle}.
The core of this principle is the mathematical property that a nonlinear differential equation of the form $dy/dx = y^q$ can be transformed into a linear equation:
\begin{equation}
    \frac{dy}{dx} = y^q \quad \Longrightarrow \quad \frac{d \ln_q y}{dx} = 1,
    \label{eq:origin_linearization}
\end{equation}
where the $q$-logarithm is defined as $\ln_q x := (x^{1-q}-1)/(1-q)$.
This $q$-logarithm serves as a natural coordinate that maps the nonlinear stationary state into a linear form.

In the TF limit, where the spatial second derivative is neglected, the quantum fluid reaches a local mechanical equilibrium.
Physically, this means that the internal generalized force (the gradient of the local chemical potential driven by interactions) must balance the external trapping force, $-\nabla V_{\text{ext}}(x)$.
Within our geometric framework, this fundamental force balance is captured by a linear response relation in the $q$-logarithmic scale:
\begin{equation}
    \frac{d}{dx} \ln_q \psi(x) = - \beta \frac{d V_{\text{ext}}(x)}{dx},
    \label{eq:linearization_principle}
\end{equation}
where $\beta$ is a positive constant related to the interaction strength and the chemical potential.

The physical origin of this linear response in the $q$-logarithmic scale is rooted in the information geometry of strongly correlated systems~\cite{Amari2016, Suyari_PartIV}.
In standard statistical mechanics, the ideal gas ($q=1$) is governed by the additive Boltzmann-Gibbs entropy, leading to a linear differential equation in the standard logarithmic scale.
However, the presence of mean-field interactions or hard-core repulsions deforms the accessible state space.
The index $q$ quantifies this geometric deformation. By transforming the system into the $q$-logarithmic coordinate, the complex nonlinear correlations are unfolded into a mathematically flat (linear) space, allowing us to describe the macroscopic density profile without relying on perturbative expansions.

\subsection{Geometric Wavefunction and Density Profile}
Integrating Eq.~(\ref{eq:linearization_principle}) yields the $q$-Gaussian effective wavefunction:
\begin{equation}
    \psi(x) \propto \left[ 1 - (1-q) \beta V_{\text{ext}}(x) \right]_+^{\frac{1}{1-q}},
    \label{eq:q_gaussian_psi}
\end{equation}
where $[X]_+ = \max(X, 0)$ denotes the cut-off, representing the physical boundary of the quantum gas cloud.
The macroscopic particle density distribution, $n(x) = |\psi(x)|^2$, is given by:
\begin{equation}
    n(x) = n(0) \left[ 1 - (1-q) \beta V_{\text{ext}}(x) \right]_+^{\frac{2}{1-q}}.
\label{eq:q_gaussian_density}
\end{equation}

\subsection{Derivation of the Standard GPE ($q=-1$)}
First, let us examine the case of $q=-1$.
Substituting $q=-1$ into Eq.~\eqref{eq:q_gaussian_density}, the density exponent evaluates to unity, yielding the profile:
\begin{equation}
    n(x) = n(0) \left[ 1 - 2 \beta V_{\text{ext}}(x) \right]_+.
\end{equation}
This is the inverted parabola profile known as the \textit{Thomas-Fermi approximation}, which describes the ground state of a standard BEC.

Furthermore, we can verify that $q=-1$ recovers the stationary Gross-Pitaevskii equation.
Using $\frac{d}{dx} \ln_{-1} \psi = \psi \frac{d\psi}{dx}$, Eq.~\eqref{eq:linearization_principle} becomes $\psi \frac{d\psi}{dx} = - \beta \frac{dV_{\text{ext}}}{dx}$.
Integrating with respect to $x$ and multiplying by $\psi$, we obtain:
\begin{equation}
    \beta V_{\text{ext}}(x) \psi + \frac{1}{2} \psi^3 = \mu' \psi,
\end{equation}
where $\mu'$ is an integration constant.
Comparing this to the stationary GPE in the TF approximation ($V_{\text{ext}}\psi + g|\psi|^2\psi = \mu\psi$), we identify the mapping between the geometric parameter $\beta$ and the physical parameters.
The chemical potential $\mu$ is uniquely determined by the normalization condition $\int |\psi(x)|^2 dx = N$ and encapsulates the microscopic interaction strength $g$.
This confirms that the Thomas-Fermi regime of the standard GPE is a mathematical consequence of the $q=-1$ geometry.


\section{Geometric Hierarchy to the Tonks-Girardeau Limit}
\label{sec:TG_limit}

Having identified $q=-1$ as the standard mean-field regime, we now explore the strong-coupling limit.

\subsection{Derivation of the Semicircle Law ($q=-3$)}
In the limit of infinite repulsion, a 1D Bose gas becomes fermionized, forming a Tonks-Girardeau (TG) gas~\cite{Girardeau1960, Kinoshita2004}.
For a system confined in a harmonic trap, $V_{\text{ext}}(x) \propto x^2$, the ground-state density profile is described by the \textit{Wigner semicircle law}:
\begin{equation}
    n_{\text{TG}}(x) = n(0) \left[ 1 - \left( \frac{x}{R} \right)^2 \right]_+^{\frac{1}{2}},
\label{eq:semicircle_law}
\end{equation}
where $R$ is the cloud radius.

Substituting $V_{\text{ext}} \propto x^2$ into our generalized geometric density, Eq.~\eqref{eq:q_gaussian_density}, the spatial dependence is proportional to $[ 1 - \text{const} \cdot x^2 ]_+^{\frac{2}{1-q}}$.
Equating the exponents ($\frac{2}{1-q} = \frac{1}{2}$) yields $q = -3$.
Thus, the strongly correlated Tonks-Girardeau regime under harmonic confinement is recovered by the geometric index $q=-3$.

\subsection{Discrete Physical Hierarchy}
\begin{figure}[htbp]
    \centering
    \includegraphics[width=\linewidth]{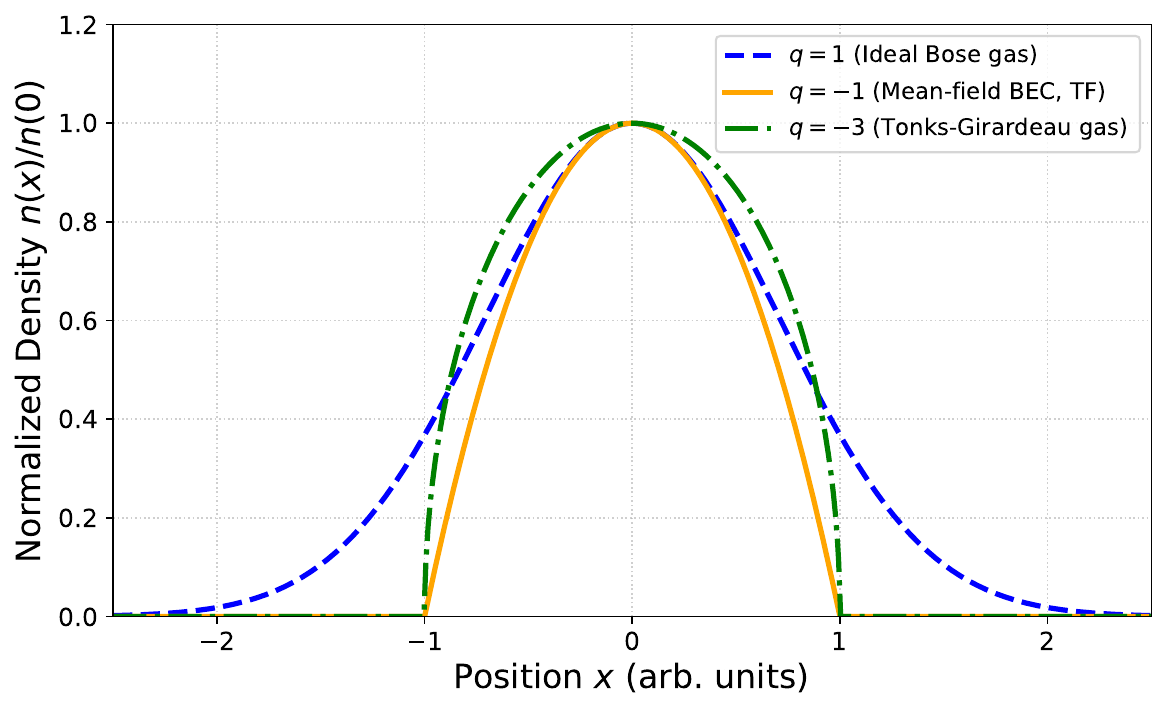}
    \caption{Normalized density profiles $n(x)/n(0)$ in a harmonic trap derived from the Linearization Principle for different values of the geometric index $q$.
The profiles demonstrate the discrete hierarchy from the ideal Bose gas ($q=1$, dashed line), to the mean-field BEC in the Thomas-Fermi limit ($q=-1$, solid line), and the strongly correlated Tonks-Girardeau gas ($q=-3$, dash-dotted line).}
    \label{fig:profiles}
\end{figure}

These results reveal a discrete hierarchy of integer $q$-values corresponding to fundamental macroscopic states (see Fig.~\ref{fig:profiles}):
\begin{itemize}
    \item $q=1$: Ideal Bose gas (Gaussian profile).
    \item $q=-1$: Standard BEC in the TF limit (Inverted parabola profile).
    \item $q=-3$: Tonks-Girardeau gas (Wigner semicircle profile).
\end{itemize}

\section{Theoretical Consistency: Connection to Nonlinear Diffusion and Geometry}
\label{sec:consistency}

In this section, we establish the thermodynamic and dynamical consistency of the geometric hierarchy by mapping our index $q$ to the polytropic index $m$ of the nonlinear diffusion equation.

\subsection{Correspondence with the Porous Medium Equation}
In the hydrodynamic limit, the macroscopic density dynamics can be modeled by the Porous Medium Equation (PME)~\cite{TsallisBukman1996}:
\begin{equation}
    \frac{\partial \rho}{\partial t} = \nabla \cdot \left[ D \nabla (\rho^m) + \rho \nabla V_{\text{ext}} \right],
    \label{eq:NFPE}
\end{equation}
whose stationary solution yields the density profile:
\begin{equation}
    \rho(x) = \rho(0) \left[ 1 - \frac{m-1}{mD} V_{\text{ext}}(x) \right]_+^{\frac{1}{m-1}}.
\label{eq:PME_solution}
\end{equation}
Comparing this spatial exponent with our geometric density exponent $2/(1-q)$, we immediately obtain the fundamental mapping $m = (3-q)/2$.

\subsection{Equations of State and Thermodynamics}
In the hydrodynamic framework, the PME index $m$ is related to the polytropic equation of state, $P \propto \rho^m$.
Applying our geometric mapping $m = (3-q)/2$ yields:
\begin{itemize}
    \item \textbf{Standard BEC ($q=-1$):} The mapping yields $m = 2$, corresponding to the equation of state $P \propto \rho^2$.
This is the pressure-density relation for a mean-field BEC arising from two-body contact interactions.
    \item \textbf{Tonks-Girardeau Gas ($q=-3$):} The mapping yields $m = 3$, corresponding to the equation of state $P \propto \rho^3$.
This recovers the degeneracy pressure exerted by a 1D gas of hard-core bosons (fermionization).
\end{itemize}
Thus, the geometric index $q$ encodes the thermodynamics of the quantum fluid across disparate interaction regimes.

\subsection{Geometric Scaling of the Sound Velocity}
The geometric framework also governs the collective dynamical excitations.
In the hydrodynamic regime, the local speed of sound $c$ is determined by the thermodynamic relation $c = \sqrt{\partial P / \partial (m_0 \rho)}$, where $m_0$ is the particle mass.
Using the geometric equation of state $P \propto \rho^{(3-q)/2}$, the sound velocity universally scales with the density as:
\begin{equation}
    c \propto \sqrt{ \frac{\partial}{\partial \rho} \left( \rho^{\frac{3-q}{2}} \right) } \propto \rho^{\frac{1-q}{4}}.
\label{eq:sound_velocity}
\end{equation}
This single geometric scaling law captures the fundamental dynamic responses of the system:
\begin{itemize}
    \item For the mean-field BEC ($q=-1$), Eq.~(\ref{eq:sound_velocity}) yields $c \propto \rho^{1/2}$, recovering the Bogoliubov sound velocity.
    \item For the Tonks-Girardeau gas ($q=-3$), it yields $c \propto \rho^1$, matching the Fermi velocity of a 1D ideal Fermi gas.
\end{itemize}

Note that this hydrodynamic scaling assumes an interaction-driven pressure. In the non-interacting limit of the ideal Bose gas ($q=1$), the Thomas-Fermi approximation breaks down as the interaction energy vanishes. Consequently, there is no macroscopic restoring force, yielding $c=0$, which is consistent with the absence of collective acoustic excitations in a non-interacting system.

Therefore, the geometric parameter $q$ not only determines the static macroscopic density profile but also dictates the dynamical response and collective excitations of the quantum many-body system in the interacting regimes.


\section{Experimental Signatures and the Crossover Regime}
\label{sec:experiment}

The analytical geometric hierarchy derived here is amenable to experimental verification using ultracold atomic gases.
By tuning the effective 1D interaction strength $\gamma$ via confinement-induced resonances, experimentalists can sweep the system across the correlation regimes.
Our framework establishes $q=-1$ and $q=-3$ as the mathematical limits (geometric fixed points) of this tunable system.
In these extreme limits, the \textit{in situ} density profiles can be observed to confirm the $q$-Gaussian forms, and the corresponding sound velocity scaling $c \propto \rho^{(1-q)/4}$ can be independently probed.

In the intermediate crossover regime ($\gamma \sim 1$), the macroscopic density profile smoothly interpolates between the inverted parabola and the Wigner semicircle, as described by the Lieb-Liniger equation of state combined with the Local Density Approximation (LDA).
Within our geometric framework, this crossover regime corresponds to a transition phase where the macroscopic state space is not globally characterized by a single integer index.
Therefore, rather than being mere asymptotic approximations, the precise integer indices $q=-1$ and $q=-3$ physically emerge as the fundamental geometric fixed points that bracket the many-body physics of 1D quantum fluids.


\section{Conclusion}
\label{sec:conclusion}

In this paper, we have proposed a geometric approach to the density profiles of 1D quantum fluids based on the \textit{Linearization Principle}.
By introducing the $q$-logarithm, we derived macroscopic stationary density profiles in the Thomas-Fermi limit without relying on perturbation theory.
We identified a discrete geometric hierarchy: the index $q=-1$ corresponds to the standard mean-field BEC (parabolic profile), whereas the index $q=-3$ yields the density profile of the Tonks-Girardeau gas (Wigner semicircle law in a harmonic trap).
We established the thermodynamic consistency of this hierarchy by mapping the geometric index $q$ to the polytropic index $m$ via the relation $m=(3-q)/2$, confirming the $P \propto \rho^2$ and $P \propto \rho^3$ equations of state.
Furthermore, this static geometric structure governs the dynamical excitations, yielding the sound velocity scaling $c \propto \rho^{(1-q)/4}$ in the interacting regimes ($q \le -1$).

The analytical geometric hierarchy is amenable to experimental verification. 
By tuning the effective 1D coupling strength, the system can be swept across the interaction regimes.
Observing the \textit{in situ} density profiles $n(x)$ and the density-dependent sound velocity provides an experimental test of the fundamental geometric boundaries, $q=-1$ and $q=-3$.

In conclusion, we have shown a geometric progression underlying the density distributions of 1D quantum fluids.
The distinct regimes of the ideal Bose gas, the mean-field condensate, and the fermionized TG gas emerge as geometric fixed points, governed by $q=1, -1$, and $-3$, respectively.
Our Linearization Principle provides an analytical framework for analyzing macroscopic states of strongly correlated quantum systems, connecting quantum mechanics, nonlinear statistical physics, and information geometry \cite{Suyari_PartIV}.


\bibliography{apssamp}

\end{document}